\documentclass[a4paper]{jpconf}
\usepackage{graphicx}
\begin{document}
\pagenumbering{arabic}
\pagestyle{plain}

\title{Review of Reactor Antineutrino Experiments}

\author{Zelimir Djurcic}

\address{Argonne National Laboratory, 9700 South Cass Avenue, Argonne, IL 60439,
 USA}

\ead{zdjurcic@hep.anl.gov}

\begin{abstract}
As discussed elsewhere, the measurement of a non-zero value for $\theta_{13}$ would open up a wide range of possibilities to explore 
CP-violation and the mass hierarchy. Experimental methods to measure currently the unknown mixing angle $\theta_{13}$ include accelerator 
searches for the $\nu_{e}$ appearance and precise measurements of reactor antineutrino disappearance.
The reactor antineutrino experiments are designed to search for a non-vanishing mixing angle $\theta_{13}$ with unprecedented sensitivity. 
This document describes current reactor antineutrino experiments and synergy between accelerator searches for the $\nu_{e}$ appearance
and precise measurements of reactor antineutrino disappearance.
\end{abstract}

\section{Introduction}
Recent years have seen enormous progress in the physics of neutrino mixing, but several critical questions remain: What is the value of $\theta_{13}$, 
the last unmeasured mixing angle in the neutrino mixing matrix? What is the mass hierarchy? Do neutrino oscillations violate CP symmetry? 
Why are the quark and neutrino mixing matrices so different? \\
Answering the questions listed above and understanding CP-violation in particular are among most important priorities within the physics community~\cite{CPref}.
A major motivation to search for CP-violation in neutrino oscillations is that its observation would make it more likely that the baryon-antibaryon asymmetry 
of the universe arose through leptogenesis~\cite{boris}.
The theory of leptogenesis is linked to the see-saw theory and as a consequence the light neutrinos are Majorana and have GUT-scale partners.
Then the matter-antimatter asymmetry of the universe may be explained through CP-violating decays of the heavy partners, producing a state with unequal 
numbers of Standard Model leptons and antileptons.
Furthermore, the Standard Model processes convert such a state into the world around us with an unequal number of baryons and antibaryons.
It is thought that CP-violation would be very unlikely to appear in the heavy sector without happening in light neutrinos~\cite{amsler}. \\
The value of $\theta_{13}$ is central to each of these questions. Measurement of non-zero value of $\theta_{13}$ would open a wide range of possibilities to explore CP-violation and the mass hierarchy. In addition, the size of $\theta_{13}$ 
with respect to the other mixing angles may give insights into the origin of these angles and the source of neutrino mass.\\
Experimental methods to measure the currently unknown mixing angle $\theta_{13}$ include accelerator searches for $\nu_{e}$ appearance
and precise measurements of reactor antineutrino disappearance.
In long-baseline appearance experiments, we search for the appearance of electron neutrinos in a fairly pure muon neutrino beam as a function of
distance $L$ from the neutrino source to the detector and the neutrino energy $E$. In such a case the neutrino oscillation probability may be 
expressed as~\cite{nunokawa}
 \begin{eqnarray}
 P(\nu_\mu \to  \nu_e) & = &
 \sin^2\theta_{23} ~\sin^2 2\theta_{13} 
~\frac{\sin^2(\Delta_{31} - aL) }
{( \Delta_{31} - aL)^2}~\Delta^2_{31}\nonumber \\
 & &+ \cos\delta ~\sin 2\theta_{23} ~ \sin 2 \theta_{12} ~\sin 2\theta_{13} ~\cos\Delta_{32} \Big( \frac{ \sin(\Delta_{31} \mp aL)}{(\Delta_{31} \mp aL)} \Delta_{31}\Big) ~\Big( \frac{ \sin(aL)}{(aL)} \Delta_{21}\Big) 
  \nonumber  \\
  & &+ \sin\delta ~\sin 2\theta_{23} ~ \sin 2 \theta_{12} ~\sin 2\theta_{13} ~\sin\Delta_{32} \Big( \frac{ \sin(\Delta_{31} \mp aL)}{(\Delta_{31} \mp aL)} \Delta_{31}\Big) ~\Big( \frac{ \sin(aL)}{(aL)} \Delta_{21}\Big) 
  \nonumber  \\
 & & +~  \cos^2\theta_{23} ~ 
\sin^2 2 \theta_{12}  ~\frac{\sin^2 ( aL)}{(aL)^2}~\Delta^2_{21}, 
\label{eq_1}
 \end{eqnarray}
 where $\Delta_{jk}$ stands for the kinematic phase, $\Delta m^2_{jk}L/4E$.
 Matter effects are described with $a \equiv G_F N_e/\sqrt{2}$ which is approximately $(3500~{km})^{-1}$. 
$\delta_{CP}$ is the CP-violating phase and the +(-) refers to antineutrino (neutrino) oscillations.
The first (last) term in Eq.(\ref{eq_1}) represents the atmospheric (solar) probabilities
 and the middle terms describe the interference between them.
It is clear that $\theta_{13}$ may be probed by measuring electron neutrino appearance from accelerator produced muon neutrinos. 
We need the values $L$ and $E$ such that interference between solar and atmospheric scales can be seen.	
Currently there are two long-baseline experiments running: the T2K~\cite{T2K} and the MINOS~\cite{minos13}.
Both recently reported indications of $\nu_\mu \to  \nu_e$ appearance.
A recent global analysis~\cite{fogli} with all available data including T2K and MINOS, provided $>$3 sigma evidence for nonzero $\theta_{13}$.
If $\theta_{13}$ is confirmed not to be too small the long-baseline oscillation experiments will have sensitivity for measuring the CP-violating phase.
However, the oscillation probability is complicated and dependent not only on $\theta_{13}$ but also on
\begin{itemize}
\item CP-violation parameter ($\delta_{CP}$),
\item Mass hierarchy	 (sign of $\Delta m^{2}_{31}$), and
\item Size of $sin^{2} \theta_{23}$.
\end{itemize}
at the same time. Therefore any attempt to measure CP-violation and the mass hierarchy would be greatly simplified if $\theta_{13}$ was measured independently. \\
A clean measurement of $\theta_{13}$ can be performed by observing the disappearance of electron antineutrinos. In general, the electron antineutrino
survival probability is given by
 \begin{eqnarray}
  P(\bar{\nu}_e \to  \bar{\nu}_e) & = &
 1 - \cos^4 \theta_{13} ~\sin^2 2\theta_{12} ~\sin^2 \Delta_{21}
 \nonumber \\
   & &- ~\sin^2 2\theta_{13} ( \cos^2 \theta_{12} ~\sin^2 \Delta_{31} +  \sin^2 \theta_{12} ~\sin^2 \Delta_{32}),
\label{eq_2}
 \end{eqnarray}
which further simplifies to
 \begin{eqnarray}
  P(\bar{\nu}_e \to  \bar{\nu}_e) & \approx &
 1 - \sin^2 2 \theta_{13}  ~\sin^2 \frac{\Delta m^2_{31} L}{4 E},
\label{eq_3}
 \end{eqnarray}
at distances 1-2 km from a reactor source. Eq. (\ref{eq_3}) shows that the reactor experiments hold the promise of unambiguously determining 
the $\theta_{13}$ mixing angle. The only additional input needed is the well measured value of $\Delta m^2_{31} \approx \Delta m^2_{atm} = (2.32^{+0.12}_{-0.08}) \times 10^{-3}~eV^2$~\cite{minos}.
As described elsewhere~\cite{bemporad}, electron antineutrinos from large commercial nuclear reactors are playing an important role 
in the exploration of neutrino oscillations. The choice of distance between source 
and detector allows one to conveniently tune an experiment's sensitivity to either the atmospheric 
or solar neutrino mass splitting.   KamLAND~\cite{kamland_prl,kamland_prl_2,kamland_zd,kamland_prl_3}
has established a connection between the MSW effect in the sun~\cite{MSW} and vacuum oscillations
with antineutrinos, and has provided the most accurate measurement of $\Delta m_{21}^2$ to date. 
Earlier, CHOOZ~\cite{chooz_last} and Palo Verde~\cite{pv_final} provided what are still the best 
upper limits on the mixing angle $\theta_{13}$.   
Antineutrino detectors at very short 
distance are being explored with the purpose of detecting coherent neutrino-nucleon 
scattering~\cite{CnuNS}, and for testing plutonium diversion at commercial 
reactors~\cite{songs}.

\section{Reactor Oscillation Searches for $\theta_{13}$}

The reactor antineutrino experiments are designed to search for a non-vanishing mixing angle $\theta_{13}$ with unprecedented sensitivity. 
The measurement of $\theta_{13}$ will be conducted by searching for an apparent disappearance of electron antineutrinos in an intense flux 
from nuclear reactor cores available to the experiments.
The current best individual constraint on the third mixing angle comes from the CHOOZ~\cite{chooz_last} reactor neutrino experiment 
$\sin^2 2 \theta_{13} < 0.15$ at 90\% C.L., with $\Delta m^2_{atm} = 2.5 \times 10^{-3}~eV^2$. 
The improvement of the CHOOZ result requires an increase in the statistics, a reduction of the systematic error below one percent, 
and a careful control of the backgrounds. 
The new generation of reactor oscillation searches will use two or more identical detectors to explore the range of  
$\sin^2 2 \theta_{13}$  down to $\sim$ 0.01.\\
Except for the case of coherent neutrino-nucleon scattering, which will not be addressed further here, 
reactor antineutrinos are usually detected using the inverse-$\beta$
reaction $\bar{\nu}_e + p \rightarrow n + e^+$ whose correlated signature helps reduce backgrounds
but is only sensitive to $\bar{\nu}_e$s with energy above 1.8 MeV. 
The kinetic energy of the positron gives a measure of the incoming $\bar{\nu}_e$ energy.
Neutrino oscillation parameters 
may be extracted from the data by fitting the observed $\bar{\nu}_{e}$ spectrum $\frac{dn_\nu}{dE_\nu}$ 
to the following equation:
\begin{equation}
\frac{dn_\nu}{dE_\nu} = \sum_k^{\textrm{reactors}} N_p \epsilon(E_\nu) \sigma(E_\nu) \frac{P_{ee}(E_\nu, L_k)}{4 \pi L_k^2} S_k(E_\nu), ~~\textrm{with} ~~S(E_\nu) = \sum_i^{\textrm{isotopes}}  f_i \left( \frac{dN_{\nu i}}{dE_\nu} \right),
\label{eq:NuebarSpecCalc}
\end{equation}
Here, $N_p$ is the number of target protons, 
$\epsilon(E_\nu)$ is the energy-dependent detection efficiency, 
$\sigma(E_\nu)$ is the detection cross-section, 
$P_{ee}(E_\nu, L_k)$ is the oscillation 
survival probability for $\bar{\nu}_e$s traveling a distance $L_k$ from reactor $k$ to the detector, 
and $S_k(E_\nu)$ is the $\bar{\nu}_e$ spectrum emitted by reactor $k$.
 $S_k(E_\nu)$ is usually expressed in form of $\frac{dN_{\nu i}}{dE_{\nu}}$ the $\bar{\nu}_e$ emission spectrum per fission of isotope $i$, 
and $f_i$ the number of fissions of isotope $i$ during the data 
taking period. Since $>$99.9\% of the energy and $\bar{\nu}_e$-producing fissions are from $^{235}$U, 
$^{238}$U, $^{239}$Pu, and $^{241}$Pu~\cite{lester_thesis}, 
the summation in Eq.(\ref{eq:NuebarSpecCalc}) is
performed over just these 4 isotopes. The time variation of the flux may be
included in the fit to discriminate the time-varying reactor signal from
constant backgrounds. 
Eq.(\ref{eq:NuebarSpecCalc}) should technically be multiplied by the
detector resolution function and integrated over $E_{\nu}$.\\
Nuclear power reactors operate on the principle that the fission of U and Pu isotopes and the 
subsequent decays of their daughter fragments release energy, generating heat.  
Large Q-value $\beta$-decays of unstable fission fragments are primarily 
responsible for the $\bar{\nu}_e$ emission of nuclear reactors.
A typical 3.8~GW reactor emits $\sim 7 \times 10^{20} ~\bar{\nu}_e/s$. Such source would   
give about 1500 $\bar{\nu}_e$ interactions per year in a 1 ton detector at 1100 m distance.
 Decays of long-lived 
isotopes in the nuclear fuel and in spent fuel elements stored at the reactor site contribute 
at the sub-percent level to the detected $\bar{\nu}_e$ flux; this contribution has
been treated elsewhere~\cite{kopeikin2001}. 
For $^{235}$U, $^{239}$Pu, and $^{241}$Pu, 
the emitted $\bar{\nu}_e$ spectra $\frac{dN_{\nu_i}}{dE_\nu}$ 
are derived from $\beta$-spectrum measurements of the fissioning of the isotopes at ILL
by thermal neutrons~\cite{schreck1985,AFF1982,hahn1989}. 
For $^{238}$U, no measurements are available, 
so theoretical calculations of its $\bar{\nu}_e$ emission must be used~\cite{vogel1981}. 
Such predictions of the antineutrino flux were considered reliable because experimental observation agreed with prediction
at 2-3\% level, thus declaring nuclear reactors as "calibrated sources" of antineutrinos~\cite{bemporad}.
A recent improved derivation antineutrino spectrum described in~\cite{huber,mueller}, based on the ILL reference $\beta$-spectrum measurements,
and updated calculation of the $^{238}$U spectrum~\cite{mueller} resulted in a $\sim$2 $\sigma$ disagreement with the data from the short-baseline
neutrino measurements~\cite{mention}. The data is lower than expectation and consistent with 
a number of hints suggesting  a possible oscillation to sterile neutrinos in the $\Delta m^2 \sim 1$ eV$^2$ region~\cite{dm1_evidence}.\\
Equally important for the understanding of the predicted antineutrino flux is a proper understanding of the systematic effects involved in the 
modeling of the reactor antineutrino flux and energy spectrum. 
The sources of systematics (thermal power, boron concentration, temperatures and densities, etc.) are usually propagated, 
through appropriate core simulation packages, to the
fission rates needed to calculate the expected antineutrino spectrum. 
Moreover, as most experiments measure $\bar{\nu}_{e}$s 
from more than one reactor, it is important to understand not only the magnitude of these effects, 
but also the correlations between uncertainties from different sources~\cite{djurcic}. 
The quality of such simulation work may be quantified through available benchmarks as shown in~\cite{c_jones} and references therein.

\section{Current Reactor Oscillation Searches}
\begin{figure}[htb!!!]
\includegraphics[angle=0, width=7.9cm, height=6.5cm]{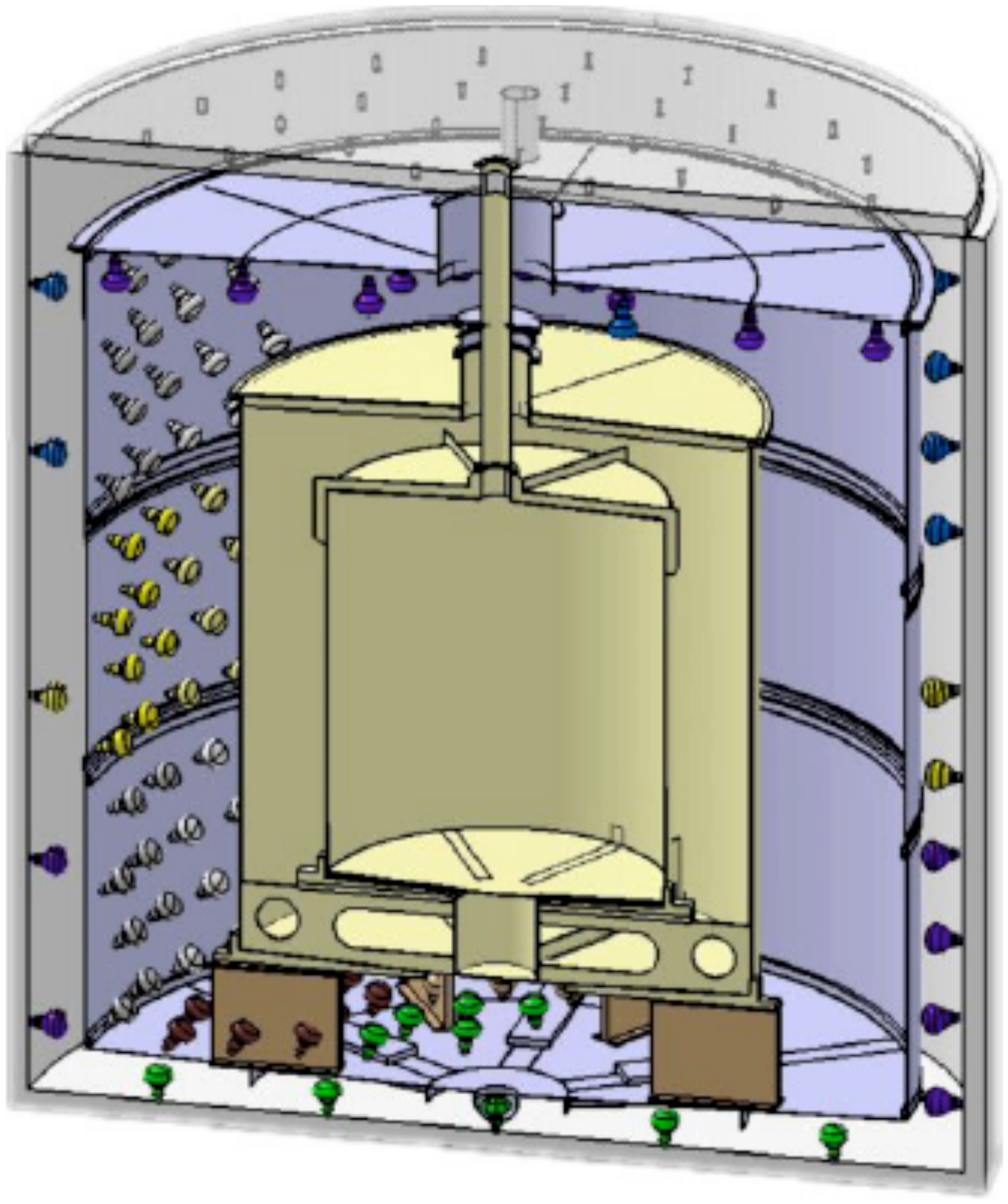} \includegraphics[width=7.9cm, height=6.5cm]{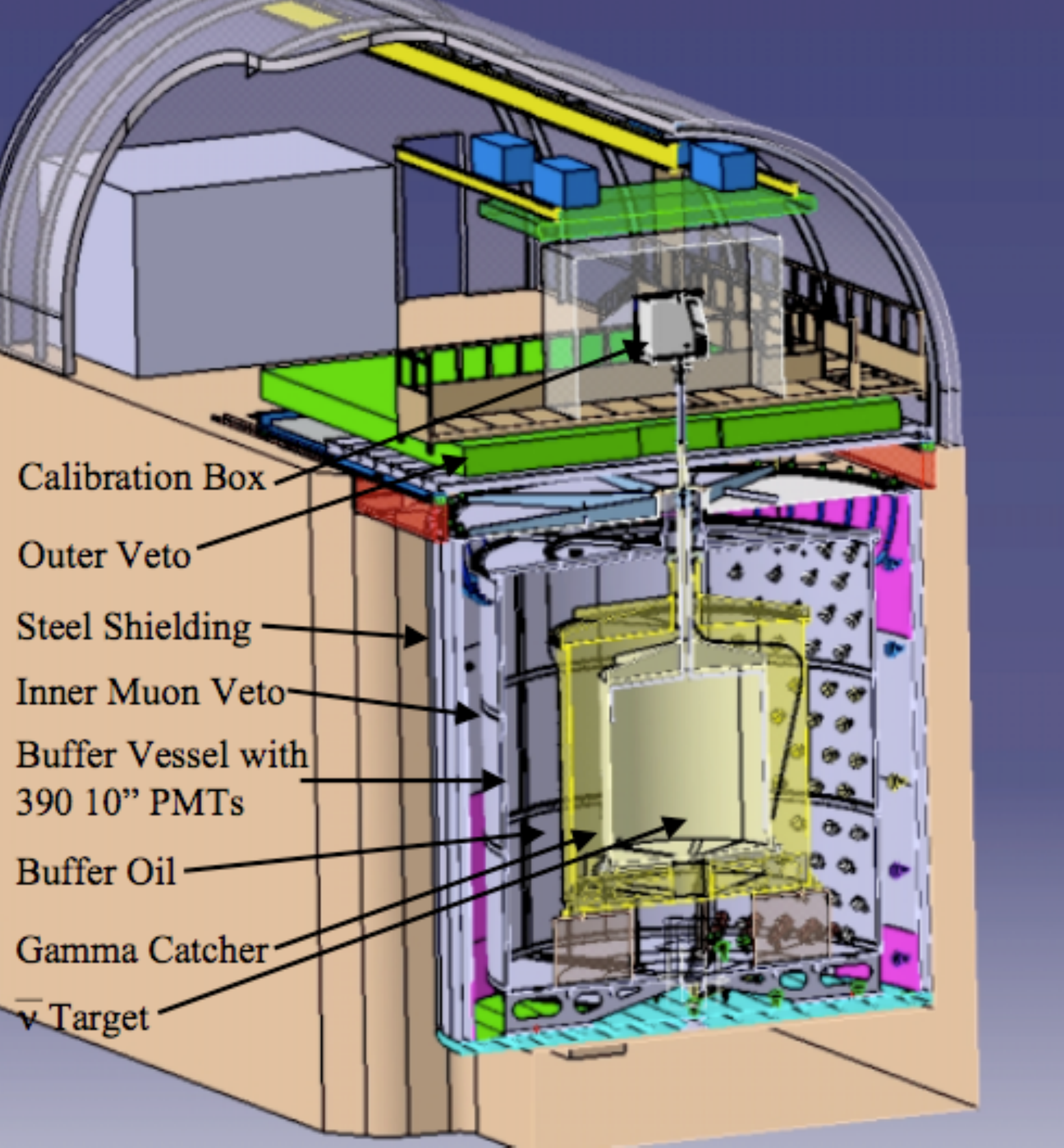}
\caption{A cross-sectional view of the Double Chooz detector. Starting from the center of the target, the detector elements are: 
Target, Gamma Catcher, Non-scintillating buffer, Buffer vessel and PMT support structure, Inner veto system, Steel Shielding, and Outer veto system. 
The detector components are enlarged on the left.}
\label{dchooz_fig}
\end{figure}

With the exception of KamLAND, long-baseline antineutrino oscillation experiment, all other current reactor experiments are short-baseline searches.
There are three reactor antineutrino disappearance experiments currently underway, constructed to perform a very sensitive measurements 
of the mixing angle $\theta_{13}$.
These include Double Chooz~\cite{dbl_chooz}, RENO~\cite{reno}, and Daya Bay~\cite{daya_bay}.
All of the current experiments use a new detector design with a gadolinium (Gd) doped liquid scintillator target surrounded by a non-doped scintillator region to detect the gammas from neutron capture on Gd. 
Antineutrinos will be detected through the inverse beta decay reaction: $\bar{\nu}_e + p \rightarrow n + e^+$. 
The $\nu_e$ target in each detector is the liquid scintillator loaded with 0.1\% Gd. Antineutrino events are identified 
by a coincidence between the prompt positron signal, which provides a direct measure of the neutrino energy, and the delayed signal from neutron capture 
on Gd.
In addition, all experiments will use a combination of near and far detectors to reduce the reactor flux uncertainty.
Compared to the original CHOOZ experiment, which used a single detector, the use of two detectors eliminates sensitivity to calculations of the reactor flux, 
and only requires a knowledge of the relative acceptance of the two detectors rather than absolute acceptance. \\
The Double Chooz experiment is currently performed at the Chooz Nuclear Power Station, the site of the original CHOOZ experiment~\cite{chooz_last}. 
The facility has two 4.27 GWth reactors separated by 140 m. Next year the experiment will be using two identical detectors: a ÒnearÓ detector at an average distance of about 400 m from the two reactor cores and a ÒfarÓ detector at an average distance of 1050 m. The experiment is currently running with the single far detector.
The far detector is located in the cavern with 300 m.w.e. overburden used by the original CHOOZ experiment. For the near detector, a tunnel is under excavation and a near lab will be equipped (120 m.w.e) by the end of 2012. \\
The RENO experiment in Korea has both the near and far detector operational from August 2011 at the Yeong Gwang six core reactor complex. The collaboration is currently taking antineutrino data with expected performance~\cite{reno_talk}. First results are expected in 2012. \\
The Daya Bay experiment will eventually have four near detectors at two near sites, and four far detectors exposed to antineutrinos from six core reactor complex. Two of the four near detectors are currently operational. The four far detectors will be deployed in 2012. Data taking with the full eight detector setup 
is expected to start in mid 2012~\cite{daya_talk}.\\
Table~\ref{table_compare2011} lists the parameters of the current generation of the short-baseline reactor antineutrino experiments.
\begin{table}[h]
\caption{\label{table_compare2011}
Comparison of the parameters of the current generation of short baseline reactor antineutrino experiments.} 
\begin{center}
\lineup
\begin{tabular}{*{7}{l}}
\br                              
\0\0Experiment & Thermal & Distance to & Shielding & Target  & Sensitivity & Status \cr 
\0\0                     &  Power &Near/Far& Near/Far  & Mass & $\sin^2 2 \theta_{13}$ & \cr
\0\0                     &  [GW$_{th}$] & [m] & [m.w.e] & [tons] & [90\%C.L.] & \cr
\mr
\0\0 Double Chooz & 8.4  & 390/1050  & 115/300 & 8/8 & 0.03  & Data taking  \cr
\0\0 (France) &   &   &   &   &   & started 2011\cr
\0\0                 &   &   &   &   &   &  far det. only\cr
\br
\0\0 RENO & 17.3  & 290/1380  & 120/450 & 16/16 & 0.02  & Data taking  \cr
\0\0 (Korea) &   &   &   &   &   & from mid-2011\cr
\br
\0\0 Daya Bay & 17.4  & 360(500)/  & 260/910 & 2x2x20 (N) & 0.01  & Data taking  \cr
\0\0 (China) &   &  1985(1615) &   & 4x20(F)  &   & started with\cr
\0\0  &   &   &   &   &   & 1 near site\cr
\br
\end{tabular}
\end{center}
\end{table}
Example of new detector design will be described in the case of the Double Chooz experiment.
The Double Chooz detector design is shown in Fig.~\ref{dchooz_fig}. It is optimized to reduce backgrounds to a negligible level. 
Each detector consists of a series of concentric cylinders. 
The target vessel is an acrylic cylinder and it contains 10.3 $m^3$ of liquid scintillator loaded with 0.1\% Gd. 
The next liquid volume is the Gamma Catcher, a 55-cm-thick buffer of non-loaded liquid scintillator with the same 
optical properties (light yield, attenuation length) as the target volume. This scintillating buffer increases the efficiency 
for detecting the photons from the neutron capture on Gd and from the positron annihilation. A 105-cm-thick region of 
non-scintillating liquid surrounds the Gamma Catcher and serves to decrease the level of accidental background. 
This buffer will reduce the singles rates in each detector by two orders of magnitude with respect to those in the
original CHOOZ experiment, which had no such buffer. A total of 390 10'' low background phototubes (PMTs) are 
mounted from the interior surface of the buffer vessel. Outside of the buffer vessel, an additional 50 cm-thick Inner veto region is filled with liquid scintillator and instrumented with PMTs. The Inner Veto is surrounded by a 15 cm Steel Shielding, protecting the detector from rock radioactivity. As a result of the few MeV energy of the reactor antineutrinos, cosmogenic backgrounds are a serious issue for reactor experiments. Double Chooz instrumented an Inner Veto 
detector in combination with the Outer Veto detector to tag cosmic-ray muons which may produce backgrounds in the target region. Outer Veto is a large coverage muon system of scintillating counters to be placed above the detector. 
It is expected to be a powerful tool in studying the $^9$Li background, which is not well understood. The combined veto system reduces backgrounds from accidentals, fast neutrons, and muon capture by large factors. With about 47,000 events expected in the far detector over a 3-year run, the statistical uncertainty associated 
with the near to far comparison is expected to be 0.5\%. In order to take advantage of this statistical precision, 
the experiment needs to reduce the systematic uncertainties to this level. 
Double Chooz will greatly reduce this systematic component by making ratio of the antineutrino spectra in two 
detectors. The antineutrino rates are proportional to the number of free protons within the target volumes. 
The relative fiducial volume uncertainty of the detectors is studied and assumed to be 0.2\%. 
Uncertainties associated with efficiencies and cuts may be reduced in 
the final analysis due to care in making the near and far detectors identical in design, materials, and target volumes. 
The Double Chooz detectors were designed to eliminate the problems with radioactivity encountered in CHOOZ by adding buffer volume 
and selecting low radioactivity materials for the detector components. Therefore, only a minimum number of analysis cuts is sufficient, 
resulting in an overall contribution to systematic uncertainty of about 0.5\% in the two-detector run.\\
A comparison of the current reactor experiments is given in Table~\ref{table_compare2011} with the available thermal power
for each experiment, the distance of near and far detectors from the reactors, shielding of detectors, detector target mass, expected sensitivities, and
current schedule.

\section{Current Status and Future Prospects}

At the time of writing this article the Double Chooz collaboration presented results of the first oscillation search performed with the far-only 
detector~\cite{dchooz_first}.
The result is an indication of reactor electron antineutrino disappearance consistent with neutrino oscillations. 
A ratio of 0.944 $\pm$ 0.016 (stat) $\pm$ 0.040 (syst) observed to predicted events was obtained in 101 days of running at 
the Chooz Nuclear Power Plant in France, with two 4.25 GW$_{th}$ reactors. The results were obtained from the far detector alone with 10m$^3$ fiducial 
volume located 1050~m from the two reactor cores. 
The Bugey4~\cite{bugey4} measurement was used to establish expectation for antineutrino rate prediction, effectively playing the role of the near detector.
The deficit can be interpreted as an indication of a non-zero value of still unmeasured neutrino mixing parameter $\theta_{13}$. Analyzing both the rate
of the prompt positrons and their energy spectrum it is found $\sin^2 2\theta_{13} = 0.086 \pm 0.041~(stat) \pm 0.030~(syst)$
corresponding to $0.015<\sin^2 2\theta_{13}<0.16$ at 90\% C.L.\\
This is one of the important first steps in our multi-year program to establish the magnitude of $\theta_{13}$ and a valuable input to today's global fits 
to the three-neutrino oscillation model. Indeed, when combined with current accelerator results~\cite{T2K}~\cite{minos13} the Double Chooz 
result leads to $\sin^2 2 \theta_{13}  \neq 0$ at 3$\sigma$ level~\cite{machado}.
Ultimately the Double Chooz will use two identical detectors to explore the range of $\sin^2 2 \theta_{13}$ from 0.15 to about 0.03, 
within three years of data taking.
RENO experiments expects first results in mid 2012 with both detectors combined.
Daya Bay experiment will be producing results soon after the full setup is operational.\\
With respect to the hints suggesting a possible oscillation to sterile neutrinos in the $\Delta m^2 \sim 1$ eV$^2$ region there are various proposals 
to perform precise searches. Considering reactor antineutrinos one may use a small cubic meter detector near research reactors and measure 
the $\bar{\nu}_e$ rate as a function of distance from the core~\cite{nucifer,scraam,danns}.\\
Aside from sterile neutrino issues and with recent results suggesting $\sin^2 2 \theta_{13} \sim 0.08$~\cite{machado}, 
the reactor experiments are in a unique
position to perform a precise measurement of this quantity. If confirmed, such a large value would open a window of opportunity for long-baseline
experiments with a rich program of neutrino oscillation physics associated with CP-violation and mass hierarchy measurements.
The upcoming NO$\nu$A~\cite{NOvA} experiment will be operational in 2014. It will be the first precision experiment to probe oscillations with
neutrinos propagating through signifiant matter (810 km distance between neutrino production point and the far detector). 
Several other long(er)-baseline experiments combined with high intensity neutrino sources are considered~\cite{shaevitz_panic11}. 
These would utilize very large 
neutrino detectors to collect sufficient event statistics at far sites. The program under development in the US is the LBNE experiment with a neutrino beam
originating at Fermilab and a detector (either water Cherenkov or liquid argon) located in proposed underground laboratory in the Homestake
mine in South Dakota~\cite{lbne}. LBNE might benefit from possible extensions such as Fermilab Project-X upgrade~\cite{projX} and/or a high-intensity 
decay-at-rest $\bar{\nu}_e$ source proposed by the DAE$\delta$ALUS collaboration~\cite{daedalus}.
Other proposals are considered worldwide including Hyper-K experiment in Japan~\cite{hyperK}, a neutrino beam from CERN to either
very large water Cherenkov detector~\cite{memphys}, liquid scintillator~\cite{lena}, or liquid argon~\cite{glacier} detector.\\
At the end, it  will be the synergy between accelerator searches for the $\nu_{e}$ appearance
and precise measurements of reactor antineutrino disappearance that helps us to measure the currently unknown mixing angle $\theta_{13}$ 
and CP-violation and mass hierarchy.

\end{document}